\documentclass[floats,twocolumn,aps,prb,longbibliography,superscriptaddress]{revtex4-2}

\usepackage{graphicx}
\usepackage{amsfonts}
\usepackage{amssymb}
\usepackage{amsbsy}
\usepackage{amsmath}
\usepackage[colorlinks = true,
            linkcolor = magenta,
            urlcolor  = magenta,
            citecolor = magenta,
            anchorcolor = magenta]{hyperref}



\begin{document}

\title{Partially disordered Heisenberg antiferromagnet with short-range stripe correlations}

\author{G. G. Blesio}
\affiliation{Jo\v{z}ef Stefan Institute, Jamova 39, SI-1000 Ljubljana, Slovenia}
\author{F. T. Lisandrini}
\affiliation{Physikalisches Institut, University of Bonn, Nussallee 12, 53115 Bonn, Germany}
\author{M. G. Gonzalez}
\affiliation{Helmholtz-Zentrum Berlin f\"ur Materialien und Energie, Hahn-Meitner Platz 1, 14109 Berlin, Germany}
\affiliation{Dahlem Center for Complex Quantum Systems and Fachbereich Physik, Freie Universität Berlin, 14195 Berlin, Germany\looseness=-1}

\begin{abstract} 
Zero-point quantum fluctuations of a N\'eel order can produce effective interactions between quasi-orphan spins weakly coupled to the lattice. On the $\sqrt{3}\times\sqrt{3}-$distorted triangular lattice, this phenomenon leads to a correlated partially disordered phase. In this article, we use matrix product state methods to study a similar model: the $S=1/2$ stuffed square lattice. Tuning the exchange amplitudes we go from a square lattice plus orphan central spins at $J'/J =0$, to the union jack lattice at $J'/J=1$, and a square lattice including all spins at $J/J'=0$. We calculate the complete antiferromagnetic phase diagram, dominated by ferrimagnetic and N\'eel orders, and compare it with existing results. Most importantly, we find a partially disordered phase in the weakly frustrated regime. In this phase, the N\'eel order from the square lattice is unaffected, while the central spins form a collective state with exponentially decaying double-striped correlations. We also study the role of quantum fluctuations by introducing an ordering staggered magnetic field on the square sublattice and find that the central spins order ferromagnetically when fluctuations from the N\'eel order are suppressed.

\end{abstract}

\maketitle 

\section{Introduction}

Quantum spin systems provide a vast playground to study all kinds of interesting phenomena. This is mainly due to the zero-point quantum fluctuations which, enhanced by magnetic frustration, lead to the emergence of novel collective many-body states. Over the last decade, a great deal of interest has been devoted to the highly-frustrated Heisenberg antiferromagnets, both theoretically and experimentally, in the search for quantum spin liquids. These highly entangled states with fractional spinon excitations have been found theoretically, for example, in two-dimensional systems such as the kagome \cite{Ran07, Yan11,Iqbal11, Lu11, Jiang12, Depenbrock12, Iqbal14, Mei17, He17, Liao17, Lauchli19, Hering19} or triangular lattice with next-nearest neighbor interactions \cite{Kaneko14, Li15, Hu15, Zhu15, Saadatmand16, Iqbal16, Wietek17, Gong17,Hu19, Gong19, Gonzalez20}. Experimentally, there are several spin-liquid candidates (or compounds that show anomalous features and are thought to lie close to one) \cite{Han12, Mingxuan15, Han16, Norman16, Khuntia20, Xing21, Wang21, Scheie21, Zeng22, Barthelemy22, Scheie22}. 

However, in some cases, fluctuations can counter-intuitively lead to ordered states. For example, classical frustrated spin systems can exhibit an accidental degeneracy in the classical limit at zero temperature, which is usually caused by a vanishing coupling energy between subsystems. This happens in the $J_1-J_2$ Heisenberg model on the square lattice at large $J_2$, where the system decouples into two independent square lattices with two independent N\'eel orders \cite{Chandra90}. However, at finite temperatures, the thermal fluctuations correlate the angle between the two subsystems, and the degeneracy is lifted. This phenomenon has been studied in many classical cases and it is usually referred to as \textit{order by disorder} \cite{Villain80, Henley89, Reimers93, Bergman07, Chern08, Mulder10}. Quantum fluctuations at zero temperature can also break degeneracies from their classical counterparts, playing a similar role as thermal fluctuations, leading to order by quantum disorder \cite{Chubukov92, Sachdev92, Lecheminant95,Bernier08, Zhitomirsky12, Savary12, Chernyshev14, Rousochatzakis15, Rau18, Schick20}.

In frustrated systems, quantum fluctuations can also prevent different subsystems from coupling. One example of this is the one-dimensionalization effect which occurs when Heisenberg chains are coupled in a frustrated manner forming a spatially anisotropic triangular lattice \cite{Zheng06, Hayashi07, Heidarian09, Starykh10, Ghorbani16, Gonzalez17, Gonzalez22}. For spin $S=1/2$ systems, even though the chains are gapless and present quasi-long range order, an inter-chain coupling of over 50$\%$ of the intra-chain coupling is needed to break the one-dimensional character \cite{Heidarian09, Ghorbani16}. For spin $S=1$ systems, an interchain coupling of about the same magnitude as the Haldane gap is needed to close it and develop a two-dimensional incommensurate spiral order \cite{Gonzalez17}. There are also several compounds in which this mechanism is thought to play a key role in the effective reduction of the dimension \cite{Kohno07, Balents10, Nilsen15, Skoulatos17, Hembacher18, Abdeldaim19}.

Recently, it was proposed that zero-point quantum fluctuations above the magnetic order can induce effective correlations between spins weakly coupled to the lattice \cite{Gonzalez19, Seifert19}. This was studied on the $\sqrt{3}\times\sqrt{3}-$distorted triangular lattice, a model proposed for the LiZn$_2$Mo$_3$O$_8$ compound, where the triangular lattice is deformed into an emergent honeycomb lattice ($J$) coupled to central spins ($J'$) \cite{Flint13}. Exact diagonalization \cite{Shimada18} and matrix product state \cite{Gonzalez19} calculations determined that for $J' > 0.2 J$ the center spins couple to the lattice, canting the N\'eel order from the honeycomb subsystem into a ferrimagnet over the whole triangular lattice. However, for $J' < 0.2 J$ the center spins remain disordered and decoupled from the lattice, forming a partially disordered phase \cite{Gonzalez19}.  

Furthermore, in the partially disordered phase, the center spins are ferromagnetically correlated at short distances. These correlations originate from a Casimir-like effect, in the sense that they are mediated by the zero-point quantum fluctuations of the N\'eel order of the honeycomb lattice. The weakly-coupled nature of the partial disorder allows to integrate out the degrees of freedom of the ordered sublattice. A second-order perturbation theory in $J'$ and $1/S$ expansions resulted in an effective model for the central spins with nearest neighbor ferromagnetic XY interactions (which dominate at large $S$) and next-nearest neighbor antiferromagnetic Ising interactions (dominant at small $S$) \cite{Seifert19}. This model exhibits a transition between a double stripe phase (dubbed $\langle 2 \rangle$) and a ferromagnetic phase at $S_c = 0.646$, where the magnetic order is destabilized by quantum fluctuations.  This is surprisingly close to $S=1/2$ and could be the mechanism responsible for the partially disordered phase. These results raise the question about the ubiquitousness of the partially disordered phase in models with a subsystem of weakly coupled spins. In other cases, the effective model governing the disordered subsystem will be different and could either lead to other exotic disordered states, or the system not hosting a partially disordered phase at all. 

\begin{figure}[!t]
    \begin{center}
        \includegraphics*[width=0.47\textwidth]{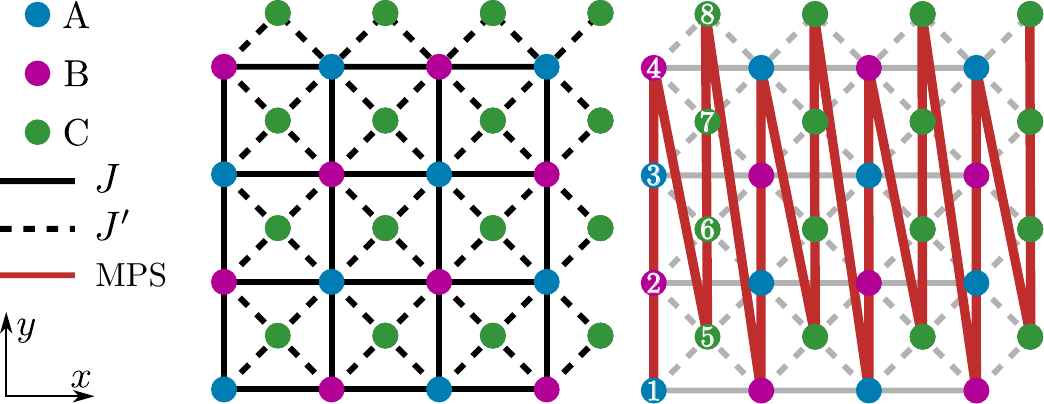}
        \caption{Stuffed square lattice where the exchange amplitude $J$ (full line) couples the spins in the square lattice and $J'$ (dashed line) couples them to the spins at the center of each square. The three sublattices A, B, and C are shown in colors. The plotted lattices are named $L_y \times L_x = 8\times4$, and the number of sites is $N = 8 \times 4 = 32$. There are as many sites C as sites AB. The red line and numbers on the right side indicate the path chosen to map the system into a one-dimensional chain for the MPS calculations.}
        \label{fig1}
    \end{center}
\end{figure}

Motivated by these questions, in this article we study the stuffed square lattice. This model is formed by a square lattice, $J$, connected to sites at the center of each square by $J'$ (see Fig.~\ref{fig1}). This allows us to study the existence of the partial disorder phenomenon and the structure of the inner correlations induced by quantum fluctuations of the N\'eel-ordered square sublattice. A similar model has been proposed in the context of the layered compound Sr$_2$TcO$_4$ compound, where frustrated center sites suppress interplane couplings \cite{Horvat2017}. 

We find that, for low values of $J'$, the spins at the center of the squares are disordered. However, their correlations present a double-stripped pattern with exponentially decaying ferromagnetic correlations along one direction and antiferromagnetic correlations to third-nearest neighbors in the perpendicular direction. In analogy with the stuffed honeycomb lattice, we argue that this is consistent with the expected effective model for the central spins.  We further study the role of quantum fluctuations in this phase by introducing an ordering magnetic field on the square sublattice. For large enough fields, quantum fluctuations are suppressed and the center sites become ferromagnetically ordered.

The rest of the article is organized as follows: in Sec.~\ref{Sec:MandM} we introduce the antiferromagnetic Heisenberg model for the stuffed square lattice and the general considerations of our matrix product states (MPS) calculations. In Sec.~\ref{Sec:Res}, we show and discuss our results, first about the general quantum phase diagram, and finally, we explore in more detail the weakly coupled limit. Finally, in Sec.~\ref{Sec:Conc} we present the summary and conclusions of our work.

\section{Model and Method}
\label{Sec:MandM}

We define the antiferromagnetic Heisenberg Hamiltonian for the stuffed square lattice as
\begin{equation}
\mathcal{H} = J \sum_{\langle i j\rangle} \mathbf{S}_i \cdot \mathbf{S}_j +  J' \sum_{\left[ i j\right]} \mathbf{S}_i \cdot \mathbf{S}_j
\label{eqH}
\end{equation}
where $J$ is the exchange interaction between nearest neighbors in the square lattice $\langle i j\rangle$ and $J'$ connects the center spins to the square lattice $\left[ i j\right]$. It is convenient to separate the lattice into three sublattices (see Fig.~\ref{fig1}). Sublattices A and B are always equivalent and form the main square lattice, and the C sublattice is composed of the spins at the center. This model has some well-known limits. For $J'/J=0$ it becomes the square lattice Heisenberg model (on the AB sublattice, with completely orphan spins C). In the opposite limit, for $J/J'=0$, the system becomes again a square lattice, but with twice the number of spins (in this limit the square lattice is formed by sublattices AB and C). Finally, the case $J=J'$ is known as the union jack lattice \cite{Collins06, Collins07, Zheng07, Bishop10, Shimokawa13, Furuya14}, a square lattice with only half of its next-nearest neighbor interactions.

To solve the model in Eq.~\ref{eqH} we use MPS methods provided by the ITensor libraries \cite{Fishman22a, Fishman22b}. In particular, we use the density-matrix renormalization group algorithm to find the ground state of the system, which requires a transformation of the two-dimensional system into a one-dimensional chain. The standard way to do this is with a zig-zag path from the bottom up and from left to right (shown by the red path in Fig.~\ref{fig1}). Two-dimensional systems are usually better represented by $L_y \times L_x$ cylinders with periodic boundary conditions along $L_y$ and $L_x \geq L_y$ \cite{White2007}. However, in our case the system is composed of two square lattices of size $L_y/2 \times L_x$, so we use $L_x \geq L_y/2$. The truncation procedure is controlled by the bond dimension, $D$. Most results are obtained with $D=3000$, while in some cases we have used up to $D=5000$ to ensure convergence of our results. Truncation errors are always kept below $10^{-6}$.

\section{Results}
\label{Sec:Res}

Throughout the article we use the following parametrization for $J$ and $J'$:
\begin{equation}
J = \cos \left(\alpha \frac{\pi}{2}\right) \qquad J' = \sin \left(\alpha \frac{\pi}{2}\right),
\end{equation}
to cover the whole range of antiferromagnetic exchange interactions with $\alpha \in [0,1]$. This way, all limits are reached: for $\alpha = 0$, $0.5$ and $1$, we get $J'/J = 0$, $1$, and $\infty$ ($J/J' = 0$), respectively. 

\subsection{Phase diagram}

\begin{figure}[!t]
    \begin{center}
        \includegraphics*[width=0.47\textwidth]{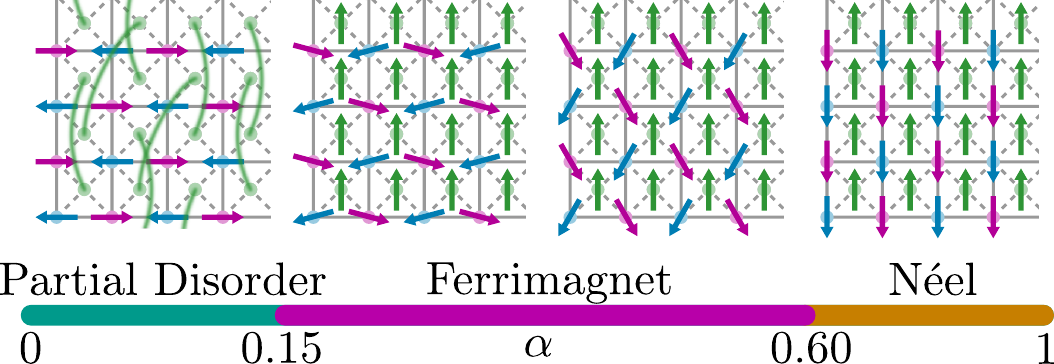}
        \caption{Schematic phase diagram of the stuffed square lattice as obtained by our MPS calculations. The phase diagram has three different phases: a correlated partially disordered phase (PD) for low values of $\alpha$, a ferrimagnetic phase (FI) for intermediate values, and a N\'eel order for large values of $\alpha$.}
        \label{fig9}
    \end{center}
\end{figure}

Before getting into the detail of our calculations, we present the phase diagram obtained, which is comprised of three distinct phases (see Fig.~\ref{fig9}): a correlated partially disordered phase (PD), a ferrimagnetically (FI) ordered phase, and a N\'eel ordered phase. For $0 \leq \alpha \lesssim 0.15 $ the system exhibits a PD phase driven by quantum fluctuations analogous to the one found in the stuffed honeycomb lattice \cite{Gonzalez19}. In this phase, AB spins form a N\'eel order with algebraically decaying correlations, while the C spins remain disordered with exponentially decaying correlations. Nonetheless, the structure of correlations is nontrivial and it is compatible with a double-striped phase. This phase, albeit ordered, was predicted for the effective model of the central spins of the stuffed honeycomb lattice \cite{Seifert19}.

For $\alpha \geq 0.15$ a first-order phase transition to a FI canted state occurs, where C spins are aligned ferromagnetically and A and B spins form an angle $\pi \pm \phi$ with C. This phase is characterized by a non-zero total magnetization. As $\alpha$ grows, the angle $\phi$ goes from $\simeq \pi/2$ to $0$, where it transitions to a N\'eel ordered phase that recovers the U(1) symmetry (see Fig.~\ref{fig9}). This transition has been studied previously, and both classical and linear spin-wave results show that $\phi = \arccos(J'/2J)$ with a critical value $\alpha_c \simeq 0.7$ ($J'=2J$) \cite{Collins06, Collins07}. On the other hand, we find $\alpha_c \simeq 0.6$, which is in close agreement with previous calculations with series expansions [$\alpha = 0.633(5)$] and coupled cluster methods [$\alpha = 0.63(1)$] \cite{Zheng07, Bishop10}. Finally, in the N\'eel phase, the spins in the AB sublattice are ferromagnetically aligned and opposite to the spins in the C sublattice.

\subsection{Ferro- and antiferromagnetic magnetizations}

An important quantity to characterize these different phases is the static spin structure factor
\begin{equation}
S^X(\mathbf{q}) = \frac{1}{N_X} \sum_{i,j \in X}\ \langle  \mathbf{S}_i \cdot \mathbf{S}_j  \rangle\ e^{i\mathbf{q} \mathbf{r}_{ij}}
\label{eq:SSF}
\end{equation}
where $X$ refers to a given sublattice (A, B, C, or AB) or the whole lattice (ABC); and $N_X$ is the number of sites in $X$. The wave-vector $\mathbf{q} = (q_x,q_y)$ is discretized along the periodic direction $q_y$, while it can take any value along $q_x$. In practice, however, both can be taken as continuous as long as we remember that important features will only appear at the values allowed by the cylindrical boundary conditions. Orders related to forbidden wave vectors are frustrated by the boundary conditions. In this sense, it is important to note that all the sublattices are square lattices (even though some are tilted to the $x$ and $y$ axes defined in Fig \ref{fig1}). We take $L_x$ and $L_y$ as even numbers to be able to host the N\'eel orders at $\alpha=0$ and 1, as well as the intermediate FI order.

As a first approach, we calculate the spin structure factor from Eq.~\ref{eq:SSF} at the points $\mathbf{q} = \mathbf{0}$ and $\mathbf{q} = {\boldsymbol \pi} = (\pm \pi,\pm \pi)$, corresponding to the ferromagnetic and antiferromagnetic N\'eel magnetizations, respectively. Explicitly, both magnetizations are written as
\begin{equation}
m_{\text{FE}}^X = \sqrt{\frac{S^X(\mathbf{0})}{N_X}} \qquad m_{\text{AF}}^X = \sqrt{\frac{S^X({\boldsymbol \pi})}{N_X}},
\label{eqmags}
\end{equation} 
where FE stands for ferromagnetic, AF for antiferromagnetic, and $X$ indicates the sublattice. It is important to say that these are not true magnetizations since in two dimensions order can only exist in the thermodynamic limit \cite{Mermin66}. A further note of caution must be added: the peak in the spin structure factor can only be associated with a semi-classical magnetization if there are no other non-equivalent peaks. We will comment when this is or is not the case. But, in general, $m$ should just be considered as a convenient measure of the signal coming from a given part of the spin structure factor.

\begin{figure}[!t]
    \begin{center}
        \includegraphics*[width=0.47\textwidth]{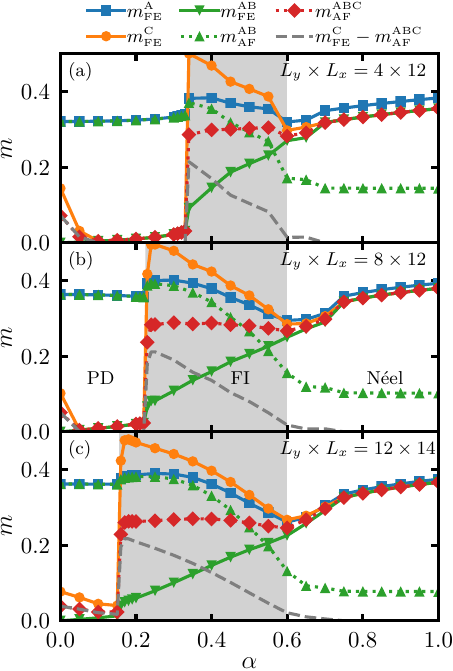}
        \caption{Ferro- and antiferromagnetic magnetizations $m_{\text{FE}}^X$ and $m_{\text{AF}}^X$ [see Eq.~\ref{eqmags} in main text] as a function of $\alpha = (2/\pi) \arctan(J'/J)$. Different sublattices are indicated in the legend, and different lattice sizes are indicated inside each panel. Results for $X=B$ are not shown since $m^B = m^A$ always. The three different phases are indicated by shaded areas and referenced in (b). The gray dashed lines corresponds to the difference $m_{\text{FE}}^C - m_{\text{AF}}^{ABC}$.}
        \label{fig2}
    \end{center}
\end{figure}

We show the values of sublattice magnetizations $m_{\text{AF}}^X$ and $m_{\text{FE}}^X$ for lattice sizes $4\times12$, $8\times12$ and $12\times 14$ in Fig.~\ref{fig2}(a), \ref{fig2}(b) and \ref{fig2}(c), respectively. The results are not sensitive to $L_x$ (not shown) and are qualitatively similar when changing $L_y$. All lattices exhibit three distinguishable phases. The PD phase at low values of $\alpha$ has the ferromagnetic magnetization from sublattices A and B equal to the antiferromagnetic magnetization from the sublattice AB, i.e. $m_{\text{FE}}^A = m_{\text{FE}}^B = m_{\text{AF}}^{AB}$. In this phase, the previous magnetizations represent the only peaks in the corresponding spin structure factors. Altogether, these features indicate that the N\'eel order from the square lattice AB at $\alpha = 0$ ($J' = 0$) extends to finite values of $\alpha$. On the other hand, the sublattice C does not have either a ferro- or antiferromagnetic magnetization, and thus it cannot be characterized by this analysis. However, it is important to note that the magnetization in the AB sublattice is unaltered by the coupling of the $C$ sublattice. That is to say, $m_{\text{AF}}^{AB}$ is almost constant in the PD phase. This also happens in the stuffed honeycomb lattice in the weakly frustrated limit \cite{Gonzalez19} and it seems to be a common feature of models where orphan spins couple in a frustrated manner to an unfrustrated lattice.

The passage to the FI phase is marked by a first-order phase transition in the C sublattice. At the phase transition, C spins develop a sudden ferromagnetic order accompanied by an almost classical value of $m_{\text{FE}}^C\simeq 0.5$ (that does not seem to decrease with the lattice size). At the same time, $m_{\text{FE}}^A = m_{\text{FE}}^B \neq m_{\text{AF}}^{AB}$, indicating that A and B sublattices no longer form a N\'eel order in AB. Instead, sublattice AB develops a small ferromagnetic magnetization that grows with $\alpha$, signaling that the spins A and B are losing the collinearity from the N\'eel order. However, sublattices A and B are still purely ferromagnetic and there is a three-sublattice order (together with C). All of these features are strongly suggesting a FI order in the whole lattice, as predicted by previous calculations \cite{Zheng07, Bishop10}. A FI phase is characterized by a total non-zero magnetization. We will use this fact to probe the FI order and calculate the angle between the three sublattices. 

Before moving on to the N\'eel phase, let us remark again the similarities to the stuffed honeycomb lattice model \cite{Gonzalez19}. First, there is a small increase of the magnetization in sublattices A and B when sublattice C orders ferromagnetically. This can be interpreted as a consequence of the reduction of quantum fluctuations caused by the sublattice C acting as a magnetic field on the sublattice AB \cite{Burkhard13, Gonzalez19}. Secondly, the magnetization sublattice C is always higher or equal to the magnetization in A and B (clearly seen in Fig.~\ref{fig2} for the $L_y=12$ lattices). This result is typical from lattices with inequivalent sites, where higher values of magnetization are observed for sites with lower effective coordination number \cite{Jagannathan06}. 

To complete the analysis of Fig.~\ref{fig2} we turn to the other end of the phase diagram (large values of $\alpha$). For the N\'eel phase, the ferromagnetic magnetizations of sublattices C and AB are the same as the antiferromagnetic magnetization on the whole lattice, $m_{\text{FE}}^C = m_{\text{FE}}^{AB} = m_{\text{AF}}^{ABC}$. This implies that the antiferromagnetic order in the whole lattice ABC is formed by two ferromagnetic sublattices C and AB (both have the same number of sites). Exactly in the limit $\alpha=1$ the system becomes a tilted square lattice with only nearest-neighbor interactions $J'$ ($J=0$). This phase extends to finite values of $J/J'$. To better visualize the phase transition, in Fig.~\ref{fig2} we plot the difference $m_{\text{FE}}^C - m_{\text{AF}}^{ABC}$ in gray dashed lines. From our calculations we can place the phase transition at $\alpha_c = 0.60(5)$, marked by the separation of the relevant magnetizations, $m_{\text{FE}}^C \neq m_{\text{AF}}^{ABC}$, i.e. $m_{\text{FE}}^C - m_{\text{AF}}^{ABC} \neq 0$. The critical point coincides with the value of $\alpha$ for which the lowest magnetization is observed for all lattices, $\alpha = 0.6$. This critical point is in close agreement with the value obtained from the series expansions and CCM calculations, $\alpha = 0.63(1)$ \cite{Zheng07, Bishop10}.

\subsection{Total magnetization and magnetic order}

\begin{figure}[!t]
    \begin{center}
        \includegraphics*[width=0.47\textwidth]{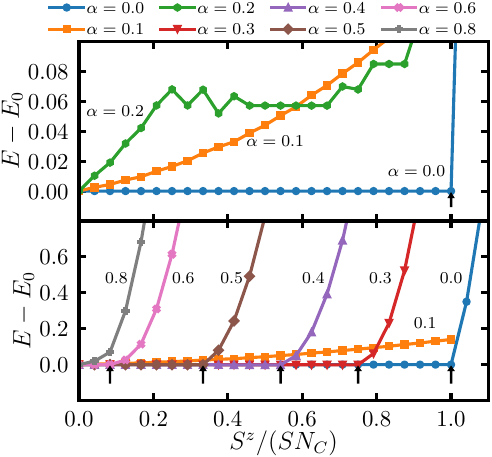}
        \caption{Energy in each subspace $S^z$ minus the ground-state energy at $S^z=0$ for different values of $\alpha$ as a function of $S^z/(SN_c)$. This normalization is chosen so that 1 indicates the fully polarized C sublattice. Both panels show the same results in different scales. The arrows indicate the maximum value of the $S^z$ subspace where the ground state can be found. Calculations correspond to the $8 \times 12$ lattice.}
        \label{fig3}
    \end{center}
\end{figure}

As mentioned above, the FI phase is characterized by a non-zero total magnetization. This means that the ground state of the system can be found in several $S^z$ subspaces. We define the subspace $S^z_\text{max}$ as the one with the largest $S^z$ that still contains the ground state. Also, $E(S^z)$ refers to the energy of the lowest-lying state in the subspace $S^z$. This implies that $E(S^z=0) = ... = E(S^z_\text{max})$ (equivalent to $E(-S^z_\text{max})$ due to inversion symmetry). In contrast, a ground state with zero total magnetization can only be found on the $S^z=0$ subspace. Therefore, $E(S^z)$ is a good quantity to differentiate the phases of our model. We show these results in Fig.~\ref{fig3}. The axis is normalized by the highest possible magnetization value of the C sublattice, $SN_\text{C}$ where $S=1/2$. Also, to be able to compare different values of $\alpha$, we always subtract the ground-state energy $E_0 = E(S^z=0)$. The results shown correspond to the $8 \times 12$ lattice, but all other lattices show equivalent behaviors. The top and bottom panels show two different energy scales to show more clearly all features.

For $\alpha = 0$, the C spins are completely decoupled and therefore do not contribute to the total energy. Thus, $E(S^z=0)= ... =E(SN_\text{C})$ and the first real excited state (hosted by the AB sublattice) can be found in the $S^z=SN_\text{C}+1$ subspace (see blue curves in Fig.~\ref{fig3}). This first excited state becomes gapless only in the thermodynamic limit, representing the magnonic excitations corresponding to the Goldstone modes. When $\alpha \neq 0$ but small, in the PD phase, we observe that the ground state has $S^z=0$ and all other subspaces represent excited states. This means that even though the C spins do not affect the magnetic order in the AB square lattice (remember that AB magnetization does not change and the order remains a N\'eel order, see Fig.~\ref{fig2}), they are not trivially decoupled as in $\alpha = 0$. Instead, they form a state which harbors the lowest energy excitations of the system. This is understandable in this regime, where $J'$ is much smaller than $J$. From these calculations, it is not possible to predict what happens to the excitations of the C sublattice in the thermodynamic limit (whether the disordered state is gapped or gapless).

For values of $\alpha$ corresponding to the FI phase, we find indeed a non-zero total magnetization. This is shown in Fig.~\ref{fig3}, where the ground-state energy is found on several consecutive subspaces. The value $S^z_\text{max}$ (indicated by black arrows) is the highest close to the transition point to the PD phase and then decreases when $\alpha$ increases approaching the N\'eel phase. At some point it becomes zero, indicating the passing to the N\'eel order in the whole lattice ABC. 

Regarding the nature of the transition between the PD and the FI phases, in Fig.~\ref{fig3} the ferromagnetic states are already visible as excited states before the transition, i.e., in the plateau at $E-E_0 \simeq 0.06$ for $\alpha=0.2$. Namely, the spin structure factor of the C spins in the plateau presents a ferromagnetic peak. These ferrimagnetic states lower their energy as $\alpha$ increases until they become the true ground state in the FI phase. The energy-level crossing along with the abrupt change in the magnetization observed in Fig.~\ref{fig2} confirms a first-order phase transition of the C spins. We note that the small deviations around the plateau come from the difficulty in obtaining well-converged results when two completely different phases have very similar energies.

\begin{figure}[!t]
    \begin{center}
        \includegraphics*[width=0.47\textwidth]{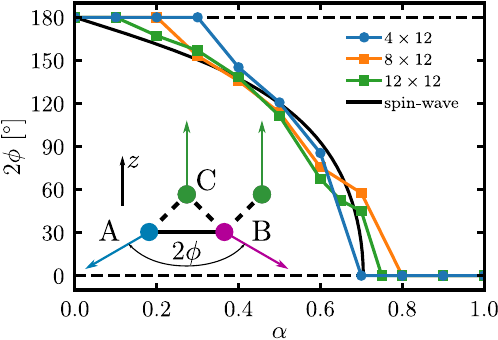}
        \caption{Semiclassical angle between spins A and B as a function of $\alpha$ for different lattice sizes [see Eq.~\ref{eq:Angle} in the main text]. The spin-wave (same as the classical) solution is shown in the black line \cite{Collins06}. The inset shows the semiclassical picture of the magnetic order.}
        \label{fig4}
    \end{center}
\end{figure}

We can further characterize the FI phase by using the semiclassical picture of the magnetic order in Ref. \cite{Gonzalez19}, shown in the inset of Fig.~\ref{fig4}. This picture assumes that all spins within a given sublattice (A, B, and C) point in the same direction with an effective size given by $m^X$ for each sublattice. This is not true for finite-size systems, which conserve SU(2) symmetry and therefore $\langle S_i^\gamma\rangle =0$ for $\gamma=x$, $y$, $z$. However, because the structure factor from each sublattice shows only a ferromagnetic peak, we can still use the semiclassical picture as an approximation.  

We know the values of $m_{\text{FE}}^X$ from the $S^z=0$ calculations shown in Fig.~\ref{fig2} (for which we recall that $m^\text{A}$ and $m^\text{B}$ are always the same). In the subspace $S^z_\text{max}$, the expectation values of the $S^z_i$ operator on each site of the sublattice C are close to the value of the magnetization $m_\text{FE}^C$ from the $S^z=0$ subspace (i.e. $\langle S_i^z\rangle \simeq m_\text{FE}^C$); and for the A and B spins, the expectation values are approximately $\langle S_i^z\rangle \simeq -m_\text{FE}^A \cos \phi$. From the total magnetization $\sum_{i\in ABC} \langle S_i^z\rangle = \sum_{i\in C} \langle S_i^z\rangle + \sum_{i\in AB} \langle S_i^z\rangle$, we then get
\begin{equation}
S^z_\text{max} = N_\text{C}\ m_\text{FE}^C - N_\text{AB}\ m_\text{FE}^A\ \cos \phi,
\end{equation}
where $2 \phi$ is the angle between spins A and B (see inset of Fig.~\ref{fig4}). Thus, we can calculate the angles in the FI phase using
\begin{equation}
2\phi = 2\arccos\left(\frac{m_\text{FE}^C - \frac{S^z_\text{max}}{N_\text{C}}}{m_\text{FE}^A}\right)
\label{eq:Angle}
\end{equation}
where we used the fact that $N_\text{AB} = N_\text{C}$ in all our lattices. It is important to note that even if this formula is strictly correct only for the FI phase, it also gives the proper results for the PD and N\'eel phases. In the N\'eel phase, all sublattices A, B, and C show only ferromagnetic peaks. In that case, we get $S^z_\text{max} = 0$ and $m_\text{FE}^C = m_\text{FE}^{AB}$ (which should replace $m_\text{FE}^A$), and thus $2\phi = 0$ indicating that the spins A and B point all in the same direction and antiparallel to those in sublattice C. Again, this is consistent with the N\'eel order over the whole lattice ABC. In the PD phase, we have to consider $m_\text{FE}^C = 0$. Since $S^z_\text{max}=0$ and $m_\text{FE}^A$ is finite, the formula results in $2\phi = \pi$, indicating that spins at A and B are antiparallel and consistent with the AB N\'eel order.

We show in Fig.~\ref{fig4} the results for the angle obtained from Eq.~\ref{eq:Angle} with the considerations mentioned above. We can see that the behavior of the angle between spins A and B is very similar to the classical and semiclassical solutions, in the sense that there is an almost linear part at low $\alpha$ to later drops abruptly at large values of $\alpha$. Close to the phase transition at $\alpha=0.6$, $\cos(\phi)$ takes values around $1$, leading to large uncertainty in the calculation of $\phi$. This can be seen in Eq.~\ref{eq:Angle} where a small deviation of $S^z_\text{max}$ from 0 leads to large changes in $2\phi$.

The key difference between our calculations and the semiclassical result is that the quantum fluctuations push the semiclassical behavior to higher values of $\alpha$ and leave a region where the AB sublattice has the same order as in $\alpha = 0$. That is, the orphan spins do not couple to the lattice and form a collective state. This effect of the zero-point quantum fluctuations is also observed in the stuffed honeycomb lattice \cite{Gonzalez19}. From our calculations, it is difficult to determine if the angle changes smoothly or not at the critical point between PD and FI phases. However, the first-order phase transition may be accompanied by a small discontinuity in the angle. As the C spins order ferromagnetically with almost classical magnetization close to the critical point and $\alpha \neq 0$, the AB spins are expected to see this change and deviate by some finite angle from the N\'eel order. 

\subsection{Structure factor and correlations}

So far we have completed the analysis and characterization of the FI phase for intermediate values of $\alpha$ and the N\'eel order for larger values of $\alpha$. We also know that at low values of $\alpha$, in the PD phase, the N\'eel order in the sublattice AB remains stable regardless of the strength with which the C spins are coupled. All it remains, then, is to characterize the state formed by the C spins in the PD phase. 

\begin{figure}[!t]
    \begin{center}
        \includegraphics*[width=0.47\textwidth]{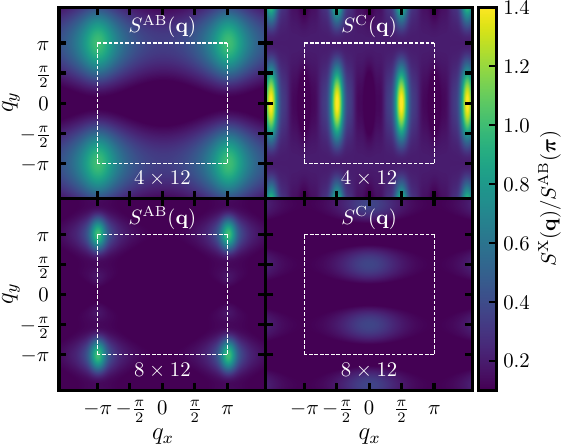}
        \caption{Spin structure factor $S^\text{X}(\mathbf{q})$ normalized by the value at the antiferromagnetic peak $S^\text{AB}(\boldsymbol{\pi})$ for $\alpha =0.1$. Lattice sizes are $4 \times 12$ and $8 \times 12$ (top and bottom, respectively), and different sublattices are AB and C (left and right, respectively). The first Brillouin zone is delimited by white dashed lines.}
        \label{fig5}
    \end{center}
\end{figure}

To do this, we start by calculating the complete spin structure factors of the AB and C sublattices, shown on the left and right panels of Fig.~\ref{fig5}, respectively. The results for the $4 \times 12$ and $8 \times 12$ lattices are shown in the top and bottom panels, respectively. For both lattices, $S(\mathbf{q})$ is normalized by the antiferromagnetic peak in the AB sublattice $S^\text{AB}(\boldsymbol{\pi})$. On one hand, the peaks in the AB sublattice appear at $S(\boldsymbol{\pi})$ and get sharper when the lattice size increases, signaling the magnetic order towards the thermodynamic limit. On the other hand, the peaks in the C sublattice appear at $(\pm \frac{\pi}{2},0)$ and $(0,\pm \frac{\pi}{2})$ in the first Brillouin zone. These peaks are consistent with a stripe pattern that repeats every four sites. Two possible orders are: a double-striped order normally dubbed $\langle 2 \rangle$ (or two-up-two-down), or a spin spiral where spins point perpendicular to the neighboring stripes. Surprisingly enough, $\langle 2 \rangle$ is one of the competing phases predicted for the effective model of C spins interacting due to quantum fluctuations of the N\'eel order in the stuffed honeycomb lattice \cite{Seifert19}. However, in the present case, the peaks get lower and diffuse when the lattice size increases, indicating a disordered state with strong quantum effects. For the $L_y = 12$ lattices (not shown), the points $(0,\pm \frac{\pi}{2})$ are not allowed in the Brillouin zone due to the cylindrical boundary conditions. Peaks at $(\pm \frac{\pi}{2},0)$ could still appear, but this does not happen. Instead, the results show that stripes tend to align along the long and non-periodic axis, but are frustrated. The next available lattice that can host this kind of peak has $L_y = 16$. For this case, we are not able to obtain quantitatively precise results and the C sublattice shows a well-established $\langle 2\rangle$ phase that spuriously breaks the SU(2) symmetry by having finite and almost classical $\langle S^z_i \rangle$. The latter decrease with increasing MPS bond dimension, but not nearly enough to obtain accurate results. This behavior of the MPS method is expected when bond dimensions are not large enough to precisely simulate the given state. 

\begin{figure}[!t]
    \begin{center}
        \includegraphics*[width=0.47\textwidth]{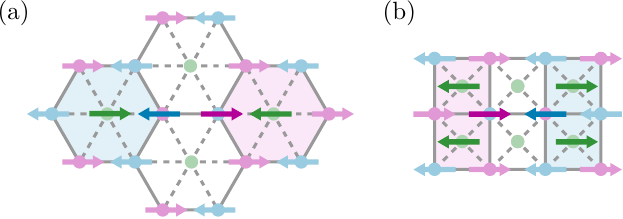}
        \caption{The effect of excitations in the AB sublattice (i.e., two-spin flip, representing a coherent two-magnon excitation) leading to effective second-nearest neighbors interactions between C spins (green). In (a), the case of the stuffed honeycomb lattice taken from Ref.~\cite{Seifert19} and in (b) the analogous excitations for the present case of the stuffed square lattice. Excited AB spins are shown in darker colors, and the effective field over C spins is shown by the shaded figures.}
        \label{fig34}
    \end{center}
\end{figure}

The similarity between the stuffed honeycomb and square lattices can be understood in terms of the same physical processes (we illustrate the similarities in Fig.~\ref{fig34}). The effective model derived in Ref. \cite{Seifert19} for the central spins has $xy$ and $zz$ interactions that depend on $S$ to several neighbours in the effective triangular lattice. For example, for $S=1/2$, antiferromagnetic $J^z$ is dominant for second nearest neighbors and decay to $J_7^z = 0.028 J_2^z$ for seventh-nearest neighbors. On the other hand, for $S=2$, ferromagnetic $J^{xy}$ is dominant to nearest neighbors, and both $|J^z|$ and $|J^{xy}|$ decay to about $1\%$ of $|J_1^{xy}|$ for fourth-nearest neighbors. There are also single-ion anisotropy terms for $S>1/2$. All in all, the $xy$ ferromagnetic interactions between C spins arise from the transverse quantum fluctuations of the ordered sublattice AB. This can be qualitatively understood considering the second-order processes involving a virtual spin-flip of one AB spin connected through the $J'$ bonds to two C spins. In our case, the same process is realized between first- and second-nearest neighbors in the square C sublattice. On the other hand, in the stuffed honeycomb lattice, antiferromagnetic Ising interactions arise from coherent two-magnon processes that connect two C spins through two AB spins [illustrated Fig.~\ref{fig34}(a)]. In our present case, these processes would lead to the same kind of interactions between third- and fourth-nearest neighbors in the square lattice C [illustrated in Fig.~\ref{fig34}(b)]. Even though this is not a formal derivation, it is to be expected that the effective Hamiltonian of the C spins, in this case, is one with $xy$ ferromagnetic interactions to first- and second-nearest neighbors and Ising antiferromagnetic interactions to third- and fourth-nearest neighbors. 

\begin{figure}[!t]
    \begin{center}
        \includegraphics*[width=0.47\textwidth]{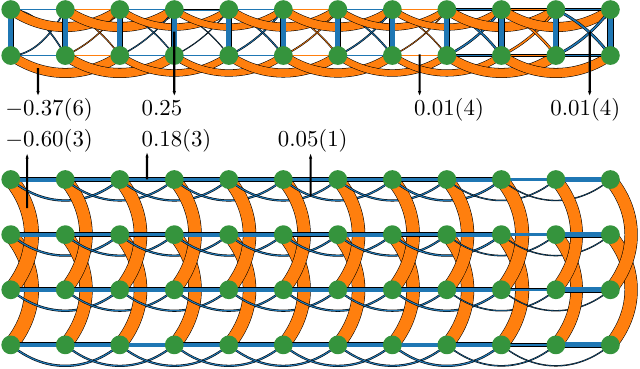}
        \caption{First- to third-nearest neighbor correlations in the C sublattice corresponding to the $4 \times 12$ (top) and $8 \times 12$ (bottom) lattices for $\alpha=0.1$. Blue and orange lines indicate ferro- and antiferromagnetic correlations $\langle  \mathbf{S}_i \cdot \mathbf{S}_j  \rangle$, respectively. The thickness is proportional to $|\langle  \mathbf{S}_i \cdot \mathbf{S}_j  \rangle|$. The numbers indicate the average and standard deviation of the type of correlation along the lattice (indicated by arrows).}
        \label{fig6}
    \end{center}
\end{figure}

To complement the spin structure factor analysis, we show in Fig.~\ref{fig6} the real-space correlations to first-, second- and third-nearest neighbors in the square C sublattice. Ferro- and antiferromagnetic correlations are indicated by blue and orange, respectively, whereas the strength is indicated by the thickness of the lines (correlations with absolute values below 0.005 are not shown for simplicity). The average values and standard deviation for each type of correlation are shown in the numbers between the two lattices. For both the $4 \times 12$ and the $8 \times 12$ lattices ($2 \times 12$ and $4 \times 12$ C sublattices), the largest correlations are ferromagnetic to nearest neighbors in one direction and antiferromagnetic to third-nearest neighbors in the perpendicular direction. This is consistent with our previous qualitative picture of effective interactions. Furthermore, as it happens for the case of the stuffed honeycomb lattice in the low-$S$ limit \cite{Seifert19}, the antiferromagnetic correlations are the dominant ones. All these features reinforce the similarities between the two lattices and strongly suggest that these effective interactions generally originate from a Casimir-like effect due to zero-point quantum fluctuations of a magnetically ordered sublattice. However, these weak effects may only be seen in these kinds of models where C spins do not interact directly with one another. As it has been suggested for the stuffed honeycomb lattice \cite{Sahoo20}, small antiferromagnetic interactions between the spins in the square sublattice C may be sufficient to order them and wash away these effects. 

\begin{figure}[!t]
    \begin{center}
        \includegraphics*[width=0.47\textwidth]{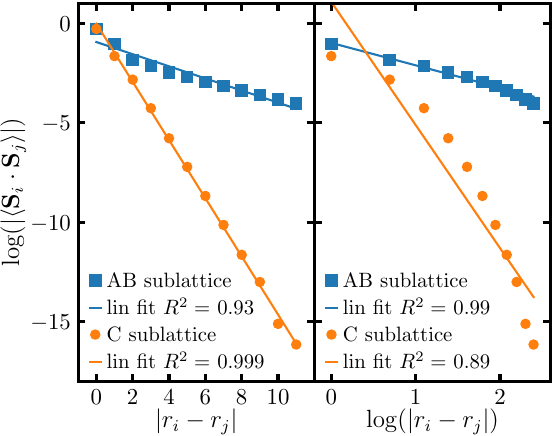}
        \caption{Correlations along the $L_x$ direction for the $8 \times 12$ lattice at $\alpha=0.1$. Blue and orange indicate the AB and C sublattices, respectively. Log-lin (left) and log-log (right) scales allow us to differentiate between exponential and power-law decaying correlations. For each case, a linear fit is performed and shown in lines of the corresponding colors.}
        \label{fig7}
    \end{center}
\end{figure}

Finally, we show in Fig.~\ref{fig7} the decay of correlations. Taking a leftmost site, we calculate all the correlations with same-type spins along the $L_x$ direction. The results do not change qualitatively if a site at the center of the system is taken, showing that the edges do not play an important role. The shown results correspond to the $8 \times 12$ lattice, and the differences with values for different $L_y$ positions are all smaller than the symbol size. The correlations used are the same on the left and right panels. On the left panel, the linear axis allows us to identify an exponential decay. The C sublattice shows a clear and strong exponential decay of the correlations through a good linear fit. This indicates short-range correlations in the C sublattice along the stripes and a gapped state. This, together with the scaling behavior of the peaks in the static structure factor, is consistent with a disordered phase. On the right panel, the logarithmic $x$ axis allows us to see power-law decaying correlations. Ordered magnets are expected to show power-law decaying correlations. In this panel we see that the AB sublattice shows good linear behavior at shorter distances, indicating an ordering tendency towards the thermodynamic limit. At large distances, the correlations deviate from the power law and agree better with exponential decay. This change of behavior is expected for MPS methods and it is connected to the finite bond dimension \cite{Schollwock2011}. The difference in the decay law of the correlations for the two sublattices is a strong indication of a PD phase.

However, the antiferromagnetic correlations between third-nearest neighbors in the perpendicular direction are strong (see Fig.~\ref{fig6}). In the $L_y=8$ lattices, it is not possible to study the decay of such correlations, because the third-nearest neighbor along $L_y$ is the same in the two directions due to the cylindrical boundary conditions. This could be the reason for such a high value of correlations, $-0.6$, which is rather close the $-0.75$ from a singlet. On the other hand, for the $L_y=4$ lattices, the strong antiferromagnetic correlations to third-nearest neighbors are along the $L_x$ direction. This lattice is too small and not very representative of the two-dimensional limit to jump to conclusions, but no decay of these correlations with the distance is observed. Also, ferromagnetic correlations along $L_y$ of exactly $0.25$ could be a signal of a triplet forming between sites. Altogether, it is difficult to determine the ground state in the thermodynamic limit from these calculations, and other methods suitable for larger lattices may be needed. However, we have collected enough evidence pointing towards a disordered but correlated state. This is consistent with a correlated PD at low values of $\alpha$. As in the case of the stuffed honeycomb lattice \cite{Gonzalez19}, short-range correlations inside the disordered sublattice are mostly ferromagnetic. But here we were able to characterize the structure of the correlations as a double stripe. 

\subsection{Stability of the partially disordered phase and classical limit}

To complete our analysis, we study the stability of the PD phase. In Ref. \cite{Seifert19}, an effective Hamiltonian for the central spins in the stuffed honeycomb lattice was derived. Such a Hamiltonian and its corresponding ground state depend on the value of the spin $S$. The authors found that for small values of $S$ the ground state is a double stripe $\langle 2 \rangle$, whereas for large $S$ the solution is ferromagnetic. Between the two phases, the fluctuations increase the corrections to the magnetization, indicating a possible disordered phase at $S_c=0.646$ (close to the quantum case $S=1/2$). Our solution for $S=1/2$ shows a disordered structure whose correlations show signatures of the $\langle 2 \rangle$ phase. Studying the same system at higher values $S$ should lead to a ferromagnetic state for the C sublattice. However, such a study is beyond the scope of this article. Instead, we approach the classical limit by applying a staggered magnetic field $h$ on the open edges of the AB sublattice, reinforcing the N\'eel order. For large enough magnetic fields, the quantum fluctuations above the N\'eel order become small and should induce a ferromagnetic in-plane order (considering a magnetic field in the $z$ direction).

\begin{figure}[!t]
    \begin{center}
        \includegraphics*[width=0.47\textwidth]{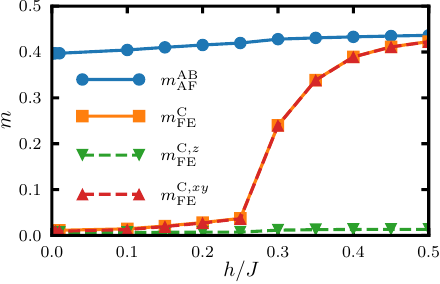}
        \caption{The antiferromagnetic magnetization $m^\text{AB}_\text{AF}$ for the AB sublattice and the ferromagnetic magnetization $m^\text{C}_\text{FE}$ for the C sublattice as a function of the staggered magnetic field $h$ applied on the AB sublattice. The solid lines correspond to the total magnetization, while the dashed lines correspond to the in-plane $xy$ and out-of-plane $z$ contributions for the C sublattice. Results correspond to the $8 \times 8$ lattice with $\alpha = 0.1$.}
        \label{fig8}
    \end{center}
\end{figure}

We show in Fig.~\ref{fig8} the magnetization as a function of the staggered magnetic field $h$ applied on the edges of the AB sublattice. The full lines show the antiferromagnetic magnetization $m^\text{AB}_\text{AF}$ for the AB sublattice and the ferromagnetic magnetization $m^\text{C}_\text{FE}$ for the C sublattice. The results shown correspond to the $8 \times 8$ lattice at $\alpha = 0.1$. As $h$ grows, the magnetization of the AB sublattice grows, getting closer to the classical value; while the ferromagnetic magnetization of C spins remains close to 0. Eventually, for $h \ge 0.3\,J$, the C spin structure factor develops a ferromagnetic peak and the ferromagnetic magnetization $m^C_\text{FE}$ grows notoriously. At $h=0$, the system preserves the SU(2) symmetry from the Hamiltonian and the in-plane contribution represents $2/3$ of the total structure factor, indicating an equal contribution of every spin direction. However, for $h\neq 0$ the Hamiltonian has only U(1) symmetry around the $z$ spin axis. For the sublattice AB, this means that the signal in the $z$ structure factor increases slowly as $h$ increases (while the signal in $xy$ decreases). On the other hand, above $h \ge 0.3\,J$, we observe that the C sublattice forms a ferromagnetic state where the in-plane contributions account for almost all the signal at the ferromagnetic peak (see dashed lines in Fig.~\ref{fig8}). This indicates that the ferromagnetic moment of the C sublattice is indeed in the $xy$-plane, perpendicular to the N\'eel order in the AB sublattice, in agreement with predictions for the effective model on the stuffed honeycomb lattice \cite{Seifert19}.

\section{Conclusions}
\label{Sec:Conc}

In this article, we have studied the stuffed square lattice, which consists of a square lattice ($J$) with extra spins at the center of each square. These central spins are connected to the spins in the square lattice by $J'$. Using a parametrization $J = \cos \left(\alpha \pi/2\right)$ and  $J' = \sin \left(\alpha \pi /2\right)$, we covered the whole range of antiferromagnetic interactions with $\alpha \in [0,1]$. Using MPS on ladders from $L_y =4$ to $12$, and $L_x$ up to $14$, we found three different phases in the corresponding quantum phase diagram. These are represented schematically in Fig.~\ref{fig9}.

For low values of $\alpha$, we found a partially disordered but correlated phase, in analogy to the previously studied stuffed honeycomb lattice \cite{Gonzalez19, Seifert19}. In this phase, the square lattice presents a N\'eel magnetic order that is unaffected by the coupling of the central spins $J'$. The central spins, on the other hand, are decoupled from the square lattice but correlated between themselves even though there is no direct exchange coupling between them. These effective correlations are induced by a Casimir-like effect, due to the zero-point quantum fluctuations of the N\'eel order in the square lattice, which acts as the medium in which the central spins are submerged. The effective Hamiltonian of the central spins seems to be dominated by short-range ferromagnetic (first and second nearest neighbors) and long-range antiferromagnetic interactions (third and fourth nearest neighbors). The latter are the dominant correlations but are not sufficient to develop a long-range magnetic order in the central-spins sublattice. Instead, a structure of double-striped correlations appears, which decay exponentially along the stripes. For finite systems, the ground state in this phase has a total $S^z=0$ with finite and very low energy excitations (of order $J'$) belonging to the central-spins sublattice. In the thermodynamic limit, the square lattice would host gapless magnon excitations corresponding to the Goldstone modes, which would be the lowest-lying excitations of the system.

When $\alpha$ grows, there is a first-order phase transition to a ferrimagnetic state. The exact point of the transition, however, seems to depend on the system size. In this sense, it is important to note that the double stripe correlations along the $x$-direction are frustrated in $L_y=12$ because of the periodic boundary conditions, causing the critical value of $\alpha$ to be particularly lower for this system size (the ferrimagnetic phase is favored). To confirm the existence of the partially disordered phase in the thermodynamic limit, we have done calculations for systems with $L_y=16$. Even though these results are not entirely reliable, we can detect a strong jump in the ferromagnetic magnetization of the sublattice C, signaling a phase transition at $\alpha_C = 0.175(25)$. All in all, we can say that the partially disordered phase should exist for $\alpha \lesssim 0.15$ in the thermodynamic limit.

For intermediate values of $\alpha$, the central spins order ferromagnetically and induce a canting angle on the square lattice. The resulting ground state is ferrimagnetic with a total spin different from 0. Contrary to most conventional ferrimagnetic systems, here the total non-zero ferromagnetic spin emerges solely from antiferromagnetic interactions in a system of equal spins. The value of $S^z_\text{max}$ is maximum close to the transition point to the partially disordered phase and decreases to $0$ as $\alpha$ grows. At $\alpha=0.6$, there is a phase transition to a N\'eel order over the whole lattice, which is a tilted square lattice dominated by $J'$. This value is in agreement with the predicted phase transition by series expansions and coupled cluster method calculations \cite{Zheng07,Bishop10}, $\alpha = 0.63$.

Finally, inspired by the calculations in Ref. \cite{Seifert19} for the effective Hamiltonian of the central spins in the stuffed honeycomb lattice, we have studied the effect of reducing the quantum fluctuations. We did so by applying a staggered magnetic field reinforcing the magnetic N\'eel order of the square lattice. We obtained that a phase transition occurs when quantum fluctuations are damped, into a phase where the central spins are ferromagnetically aligned. Furthermore, the ferromagnetic C correlations exist only in the plane perpendicular to the N\'eel order in the AB sublattice. This result is consistent with the large-S results of the effective Hamiltonian for the stuffed honeycomb lattice.

The most interesting part of the phase diagram lies in the weakly-frustrated regime at low values of $\alpha$ or $J'/J$, as it is the only phase that does not appear in the semiclassical solution of the model. And it is, therefore, of purely quantum origin. This adds up to the previous calculations on the stuffed honeycomb lattice and shows that the partially disordered phase is common in these kinds of systems. Also, the disordered sublattice is not trivially disordered and the inner correlations of the central spins develop a double stripe pattern, originated from the quantum fluctuations from the ordered sublattice. In conclusion, these kinds of systems with weakly coupled spins to a magnetically ordered lattice provide an interesting playground to study exotic quantum states.

\section*{Acknowledgments}

The authors thank L. O. Manuel, U. F. P. Seifert, and J. Reuther for the fruitful discussions. G.~G.~B. is supported by Slovenian Research Agency (ARRS) under Grant no. P1-0044 and J1-2458. Part of the computation was performed on the supercomputer Vega at the Institute of Information Science (IZUM) in Maribor, Slovenia. M.~G.~G. acknowledges usage of the JUWELS cluster at the Forschungszentrum J\"ulich, Germany. F.~T.~L. acknowledges funding from the Deutsche Forschungsgemeinschaft (DFG, German Research Foundation) in particular under Germany’s Excellence Strategy – Cluster of Excellence Matter and Light for Quantum Computing (ML4Q) EXC 2004/1 – 390534769.

\bibliography{papers}

\end{document}